\documentclass[12pt]{article}

\usepackage{amssymb}
\usepackage{amsmath}
\usepackage{amsfonts}
\usepackage{amsbsy}

\begin{document}

\title{General Slow-Roll Spectrum for Gravitational Waves}

\author{
Jinn-Ouk Gong\footnote{jgong@muon.kaist.ac.kr} \\
{\em Department of Physics, KAIST, Daejeon, Republic of Korea}}

\date{\today}

\maketitle

\begin{abstract}

We derive the power spectrum $\mathcal{P}_\psi(k)$ of the gravitational waves
produced during general classes of inflation with second order corrections.
Using this result, we also derive the spectrum and the spectral index in the
standard slow-roll approximation with new higher order corrections.

\end{abstract}

\vspace*{-65ex}

\hspace*{\fill} KAIST-TH/2004-10

\thispagestyle{empty}
\setcounter{page}{0}
\newpage
\setcounter{page}{1}

\section{Introduction}

Inflation \cite{inf}, among many of its features, is believed to have created
the scalar perturbations from primordial quantum fluctuations in the inflaton
field $\phi$. However, not only scalar curvature perturbations are produced
during inflation. Tensor fluctuations associated with the metric are also
generated along with the scalar perturbations \cite{gw}. The tensor
fluctuations, which emerge as gravitational waves, are not accompanied with
density perturbations responsible for the structure formation in the observed
universe. This makes them not directly accessible to observations. They are
currently believed to just influence the cosmic microwave background
\cite{wmap} on large angular scales and the polarization \cite{pol}, which is
one of the major aims for the future cosmic microwave background observations.
Gravitational wave detectors such as the Laser Interferometer Space Antenna
(LISA) \cite{lisa} and the Laser Interferometric Gravitational Wave Observatory
(LIGO) \cite{ligo} are currently in progress as well.

Gravitational waves, although not observed yet, possess several potential uses.
One noteworthy feature is that they are directly associated with the energy
scale of inflation. Once detected, they lead us to determine the inflationary
energy scale without any obstacle. Moreover, for the case of models of inflation
with a single degree of freedom for $\phi$, the tilt of the tensor perturbations
is related to the tensor-to-scalar amplitude ratio, known as the consistency
relation. If this relation turns out to be false, we should abandon the single
field inflationary models and consider the models with multiple degrees of
freedom for $\phi$ \cite{multi,gs} or the models in generalised gravity
\cite{ggr}. Also, it is interesting that different models for the generation of
the universe, such as ekpyrotic \cite{ekpyrosis}, cyclic \cite{cyclic} and
pre-big bang models \cite{prebb}, generally predict the power spectrum of
gravitational waves strongly tilted to the blue. Therefore detection of the
tensor perturbations would be a powerful discriminator for these models, as well
as inflation.

In this paper, we follow the Green's function method \cite{sg,gs} within the
generalised slow-roll approximation \cite{gsr,cgs} and present the power
spectrum for the tensor perturbations, $\mathcal{P}_\psi(k)$, with second order
corrections. Throughout this paper, we set $c = \hbar = 8\pi G = 1$.

\section{Power spectrum for gravitational waves}

In this section, we derive the power spectrum for gravitational waves produced
during inflation. First, we present the basic principles, and then derive the
formulae for $\mathcal{P}_\psi(k)$ in the general slow-roll scheme.

\subsection{Preliminaries}

The linear tensor perturbations in general can be written as
\cite{metric}
\begin{equation}
ds^2 = a^2 (\eta) \left[ d\eta^2 - \left( \delta_{ij} + 2 h_{ij}
\right) dx^i dx^j \right],
\end{equation}
where we assume that $|h_{ij}| \ll 1$. Although the tensor $h_{ij}$ has 6
degrees of freedom, imposing traceless and transverse conditions, we can remove
4 degrees of freedom and are left with 2 physical degrees of freedom, or
polarizations. Thus, we can write the tensor $h_{ij}$ in Fourier components as
\begin{equation}
h_{ij} = \int \frac{d^3 k}{(2\pi)^{3/2}} \sum_{\lambda=1}^2
\psi_{\mathbf{k},\lambda}(\eta) e_{ij}(\mathbf{k},\lambda) e^{i
\mathbf{k \cdot x}}
\end{equation}
where the polarization tensors $e_{ij}$ satisfy the relations
\begin{equation}
e_{ij} = e_{ji}, \ \ \ e^i_i = 0, \ \ \ k^i e_{ij} = 0, \nonumber
\end{equation}
\begin{equation}
e_{ij}^*(\mathbf{k},\mu) e^{ij}(\mathbf{k},\nu) = \delta_{\mu\nu}
\hspace{1cm} \mbox{and} \hspace{1cm} e_{ij}(-\mathbf{k},\lambda) =
e_{ij}^*(\mathbf{k},\lambda),
\end{equation}
and the power spectrum for the tensor perturbations, or gravitational waves, is
defined as
\begin{equation}
\label{psoriginal}
\langle \psi_{\mathbf{k},\lambda} \psi_{\mathbf{l},\lambda}^*
\rangle \equiv \frac{2\pi^2}{k^3} \mathcal{P}_\psi(k) \delta^{(3)}
(\mathbf{k - l}).
\end{equation}

\subsection{General slow-roll formulae for $\mathcal{P}_\psi(k)$}

To apply the general slow-roll scheme, we begin with the action for the tensor
perturbations \cite{paction}
\begin{eqnarray}\label{action}
S & = & \int \frac{1}{2} \, a^2 \left[ \left( \frac{\partial
h_{ij}}{\partial\eta} \right)^2 - \left( \nabla h_{ij} \right)^2
\right] d\eta\,d^3x \nonumber
\\ & = &
\int \frac{1}{2} \sum_{\lambda=1}^2 \left[ \left| \frac{\partial
v_{\mathbf{k},\lambda}}{\partial\eta} \right|^2 - \left( k^2 -
\frac{1}{a} \frac{d^2 a}{d\eta^2} \right) \left|
v_{\mathbf{k},\lambda} \right|^2 \right] d\eta \, d^3k \, ,
\end{eqnarray}
where we define
\begin{equation}
v_{\mathbf{k}, \lambda} \equiv a \psi_{\mathbf{k}, \lambda}.
\end{equation}
Then, the equation of motion for the Fourier modes is given as
\begin{equation}
\label{eomv}
\frac{d^2 v_k}{d\eta^2} + \left( k^2 - \frac{1}{a}\frac{d^2
a}{d\eta^2} \right) v_k = 0,
\end{equation}
where the solution $v_k$ satisfies the boundary conditions
\begin{equation}
v_k \longrightarrow \left\{ \begin{array}{ccc}
\frac{1}{\sqrt{2k}\,} e^{-ik\eta} & \mbox{as} & -k\eta \rightarrow
\infty
\\
A_k a & \mbox{as} & -k\eta \rightarrow 0.
\end{array}\right.
\end{equation}
Now, defining $\upsilon = \sqrt{2k} v_k$, $x = -k \eta$ and
\begin{equation}
\label{p}
p(\ln x) = \frac{2\pi}{k}xa,
\end{equation}
Eq.~(\ref{eomv}) becomes
\begin{equation}
\label{eomomega} \frac{d^2\upsilon}{dx^2} + \left( 1 -
\frac{2}{x^2} \right) \upsilon = \frac{1}{x^2} \, q \, \upsilon,
\end{equation}
where
\begin{equation}
q = \frac{p'' - 3 p'}{p}
\end{equation}
and $p' \equiv dp/d\ln x$. Then, we can write the solution of
Eq.~(\ref{eomomega}) using Green's function as
\begin{eqnarray}\label{intsol}
\upsilon(x) & = & \upsilon_0(x) + \frac{i}{2} \int_x^\infty
\frac{du}{u^2} q(u) \left[ \upsilon_0^*(u) \upsilon_0(x) -
\upsilon_0^*(x) \upsilon_0(u) \right] \upsilon(u) \nonumber
\\ & \equiv &
\upsilon_0(x) + L(x,u) \upsilon(u),
\end{eqnarray}
where
\begin{equation}
\upsilon_0(x) = \left( 1 + \frac{i}{x} \right) e^{ix}
\end{equation}
is the desired homogeneous solution of Eq.~(\ref{eomomega}). The power spectrum
for the tensor perturbations, $\mathcal{P}_\psi(k)$, can be written, using
Eqs.~(\ref{psoriginal}) and (\ref{p}), as
\begin{eqnarray}
\label{beforeps}
\mathcal{P}_\psi(k) & = & \frac{k^3}{2\pi^2} \lim_{-k\eta
\rightarrow 0} \left| \frac{v_k}{a} \right|^2 \nonumber
\\ & = &
\lim_{x \rightarrow 0} \left| \frac{x\upsilon}{p} \right|^2.
\end{eqnarray}

Here, we note that Eq.~(\ref{eomv}) has exactly the same structure as that for
the scalar perturbations, which was the subject of Ref.~\cite{cgs}. The only
difference is the definition of the function $p(\ln x)$ in Eq.~(\ref{p}). So,
we can just follow the same steps as those used in Ref.~\cite{cgs} to write
$\mathcal{P}_\psi(k)$. We assume that $\upsilon(x)$ is dominated by the scale
invariant, homogeneous solution $\upsilon_0(x)$, that is, $q$ is small. Then,
for the spectrum with second order corrections, we iterate Eq.~(\ref{intsol})
twice,
\begin{equation}
\upsilon(x) \simeq \upsilon_0(x) + L(x,u) \upsilon_0(u) + L(x,u) L(u,v)
\upsilon_0(v),
\end{equation}
and substituting into Eq.~(\ref{beforeps}), after some calculations, we can
obtain $\mathcal{P}_\psi(k)$ as
\begin{eqnarray}
\label{ps}
\ln \mathcal{P}_\psi(k) & = & \ln \left(\frac{1}{p_\star^2}\right) - 2
\int_0^\infty \frac{du}{u} \, w_\theta(x_\star,u) \, \frac{p'}{p} + 2 \left[
\int_0^\infty \frac{du}{u} \, \chi(u) \, \frac{p'}{p} \right]^2 \nonumber
\\ & &
\mbox{} - 4 \int_0^\infty \frac{du}{u} \, \chi(u) \, \frac{p'}{p}
\int_u^\infty \frac{dv}{v^2} \, \frac{p'}{p} \, ,
\end{eqnarray}
where the subscript $\star$ means some convenient point of evaluation around
horizon crossing, and the window functions are given by
\begin{equation}
w(x) = \frac{\sin(2x)}{x} - \cos(2x), \ \ \ \chi(x) = \frac{1}{x} -
\frac{\cos(2x)}{x} - \sin(2x),
\end{equation}
and
\begin{equation}
w_\theta(x_\star,x) = w(x) - \theta(x_\star-x).
\end{equation}
We can see that the integrals in Eq.~(\ref{ps}) are well defined as $x
\rightarrow 0$, since $w(x)$ and $\chi(x)$ behave asymptotically as
\begin{equation}
\lim_{x \rightarrow 0} w(x) = 1 + \mathcal{O}\left(x^2\right) \ \
\ \mbox{and} \ \ \ \lim_{x \rightarrow 0} \chi(x) = \frac{2}{3}
x^3 + \mathcal{O}\left(x^5\right).
\end{equation}
It is crucial to realise that Eq.~(\ref{ps}) is independent of the evaluation
point $\star$, i.e. we have the same spectrum irrespective of when we evaluate.
It is manifest from Eq.~(\ref{ps}) that the $\star$ dependence cancels out
because of the step function. Note that although the form of Eq.~(\ref{ps}) is
the same as that for the scalar perturbations \cite{cgs}, differences appear as
we specify the function $p(\ln x) = \frac{2\pi}{k}xa$, compared with $f(\ln x)
= \frac{2\pi}{k}xa\frac{\dot\phi}{H}$ for the scalar case. Some examples of
different $p$'s will be presented in the following sections.

\section{Applications}

In this section, using our result, Eq.~(\ref{ps}), we present
$\mathcal{P}_\psi(k)$ for several physically interesting cases. The utility of
this section is twofold : first, it gives specific examples of our result, so
that we can see the broad applicability of Eq.~(\ref{ps}). Also, we can verify
that it successfully reproduces the familiar, well known result for the spectrum
\cite{gw}
\begin{equation}
\mathcal{P}_\psi(k) = \left( \frac{H}{2\pi} \right)^2
\end{equation}
and the additional corrections which make $\mathcal{P}_\psi(k)$ more accurate.

\subsection{Standard slow-roll approximation}

It is very illuminating to use Eq.~(\ref{ps}) to derive $\mathcal{P}_\psi(k)$
in the standard slow-roll approximation with one higher order, i.e. third order
corrections, since we have all the necessary information with only one
undetermined coefficient \cite{cgs}. Previously, $\mathcal{P}_\psi(k)$ was
known up to second order corrections in the standard slow-roll approximation
\cite{llms,st} and third order corrections under some special conditions
\cite{st}. Here we give the third order corrections in the standard slow-roll
approximation.

From Eq.~(\ref{ps}), we can remove the logarithm and expand $p'/p$ in terms of
$\ln(x/x_\star)$, which implies we are applying the standard slow-roll
approximation, to obtain the result
\begin{eqnarray}
\label{psnotlog}
\mathcal{P}_\psi(k) & = & \frac{1}{p_\star^2} \left\{ 1 -
2\alpha_\star \frac{p'_\star}{p_\star} + \left( -\alpha_\star^2 +
\frac{\pi^2}{12} \right) \frac{p''_\star}{p_\star} + \left(
3\alpha_\star^2 - 4 + \frac{5\pi^2}{12} \right) \left(
\frac{p'_\star}{p_\star} \right)^2 \right. \nonumber
\\ & &
\hspace{3em} \mbox{} + \left[ - \frac{1}{3}\alpha_\star^3 +
\frac{\pi^2}{12}\alpha_\star - \frac{4}{3} + \frac{2}{3}\zeta(3)
\right] \frac{p_\star'''}{p_\star} \nonumber
\\ & &
\hspace{3em} \mbox{} + \left[ 3\alpha_\star^3 - 8\alpha_\star +
\frac{7}{12}\pi^2\alpha_\star + 4 - 2\zeta(3) \right]
\frac{p_\star'p_\star''}{p_\star^2} \nonumber
\\ & &
\left. \hspace{3em} \mbox{} + \left[ -4\alpha_\star^3 +
16\alpha_\star - \frac{5}{3}\pi^2\alpha_\star - 8 + 6\zeta(3)
\right] \left( \frac{p_\star'}{p_\star} \right)^3 + 
\cdots \right\},
\end{eqnarray}
where
\begin{equation}
\alpha_\star \equiv \alpha - \ln x_\star, \ \ \ \alpha \equiv 2 -
\ln 2 - \gamma \simeq 0.729637,
\end{equation}
$\gamma \simeq 0.577216$ is the Euler-Mascheroni constant, $\zeta$ is the
Riemann zeta function and we have used the results of Ref.~\cite{cgs}, where we
determined the last coefficient using some exactly known solutions.

Now, we write the slow-roll parameters in the standard slow-roll
approximation\footnote{Note that these parameters are defined differently than
in Refs.~\cite{llms,st}, where $\epsilon_0 = -\frac{\dot H}{H^2}$ and
$\epsilon_{n+1} \equiv \frac{d\ln|\epsilon_n|}{dN}$.}
\begin{equation}
\label{srp}
\epsilon_1 = -\frac{\dot H}{H^2} = \mathcal{O}(\xi), \ \ \
\mbox{and} \ \ \ \epsilon_{n+1} = \frac{1}{H^n} \left(
\frac{d}{dt} \right)^n \epsilon_1 = \mathcal{O}(\xi^{n+1}),
\end{equation}
where $\xi$ is some small parameter and $n \geq 1$ is some integer. Using these
parameters, we can write the function $p$ in terms of these slow-roll parameters
as
\begin{eqnarray}
\frac{1}{p^2} & = & \left( \frac{H}{2\pi} \right)^2 \left[ 1 - 2\epsilon_1 +
\epsilon_1^2 - 2\epsilon_2 - 2\epsilon_1\epsilon_2 - 2\epsilon_3 +
\mathcal{O}(\xi^4) \right], \nonumber
\\
\frac{p'}{p} & = & -\epsilon_1 - \epsilon_1^2 - \epsilon_2 - \epsilon_1^3 -
4\epsilon_1\epsilon_2 - \epsilon_3 + \mathcal{O}(\xi^4) , \nonumber
\\
\frac{p''}{p} & = & \epsilon_1^2 + \epsilon_2 + 2\epsilon_1^3 +
6\epsilon_1\epsilon_2 + \epsilon_3 + \mathcal{O}(\xi^4) , \nonumber
\\
\frac{p'''}{p} & = & -\epsilon_1^3 - 4\epsilon_1\epsilon_2 - \epsilon_3 +
\mathcal{O}(\xi^4) .
\end{eqnarray}
Substituting these into Eq.~(\ref{psnotlog}), we obtain the power spectrum as
\begin{eqnarray}
\label{psssr} \mathcal{P}_\psi(k) & = & \left(
\frac{H_\star}{2\pi} \right)^2 \left\{ 1 + \left( 2\alpha_\star -
2 \right) \epsilon_{1\star} + \left( 2\alpha_\star^2 -
2\alpha_\star - 3 + \frac{\pi^2}{2} \right) \epsilon_{1\star}^2 +
\left( -\alpha_\star^2 + 2\alpha_\star - 2 + \frac{\pi^2}{12}
\right) \epsilon_{2\star} \right. \nonumber
\\ & &
\hspace{2cm} + \left[ \frac{4}{3}\alpha_\star^3 - 8\alpha_\star +
\pi^2\alpha_\star + \frac{16}{3} - \frac{14}{3}\zeta(3) \right]
\epsilon_{1\star}^3 \nonumber
\\ & &
\hspace{2cm} + \left[ -\frac{5}{3}\alpha_\star^3 + 2\alpha_\star^2
+ 8\alpha_\star - \frac{11\pi^2}{12}\alpha_\star - \frac{26}{3} +
\frac{7\pi^2}{6} - \frac{2}{3}\zeta(3) \right]
\epsilon_{1\star}\epsilon_{2\star} \nonumber
\\ & &
\hspace{2cm} \left. + \left[ \frac{1}{3}\alpha_\star^3 - \alpha_\star^2 +
2\alpha_\star - \frac{\pi^2}{12}\alpha_\star - \frac{2}{3} + \frac{\pi^2}{12} -
\frac{2}{3}\zeta(3) \right] \epsilon_{3\star} \right\}.
\end{eqnarray}
In addition, we can calculate the spectral index
\begin{equation}
n_\psi(k) \equiv \frac{d\ln\mathcal{P}_\psi(k)}{d\ln k}
\end{equation}
easily from $\mathcal{P}_\psi(k)$ as
\begin{eqnarray}
n_\psi(k) & = & -2\epsilon_{1\star} - 2\epsilon_{1\star}^2 + \left(
2\alpha_\star - 2 \right) \epsilon_{2\star} \nonumber
\\ & &
- 2\epsilon_{1\star}^3 + \left( 6\alpha_\star - 12 + \pi^2 \right)
\epsilon_{1\star}\epsilon_{2\star} + \left( -\alpha_\star^2 +
2\alpha_\star - 2 + \frac{\pi^2}{12} \right) \epsilon_{3\star}
\nonumber
\\ & &
-2\epsilon_{1\star}^4 + \left[ -\frac{5}{3}\alpha_\star^3 + 2\alpha_\star^2 +
20\alpha_\star - \frac{11\pi^2}{12}\alpha_\star - \frac{92}{3} +
\frac{31\pi^2}{6} - \frac{44}{3}\zeta(3) \right]
\epsilon_{1\star}^2\epsilon_{2\star} \nonumber
\\ & &
+ \left[ \alpha_\star^3 - 7\alpha_\star^2 + 22\alpha_\star -
\frac{5\pi^2}{4}\alpha_\star - 16 + \frac{19\pi^2}{12} - 2\zeta(3) \right]
\epsilon_{1\star}\epsilon_{3\star} \nonumber
\\ & &
+ \left[ \frac{1}{3}\alpha_\star^3 - 4\alpha_\star^2 + 16\alpha_\star -
\frac{13\pi^2}{12}\alpha_\star - \frac{38}{3} + \frac{4\pi^2}{3} -
\frac{2}{3}\zeta(3) \right] \epsilon_{2\star}^2 \nonumber
\\ & &
+ \left[ \frac{1}{3}\alpha_\star^3 - \alpha_\star^2 + 2\alpha_\star -
\frac{\pi^2}{12}\alpha_\star - \frac{2}{3} + \frac{\pi^2}{12} -
\frac{2}{3}\zeta(3) \right] \epsilon_{4\star}.
\end{eqnarray}
Note that the slow-roll parameters, Eqs.~(\ref{srp}), are given as functions of
$H$ only, so these results are applicable to models with multiple degrees of
freedom for inflaton field $\phi$ \cite{multi,gs}, as well as single field cases
\cite{sg,gsr,cgs}.

\subsection{de Sitter background}

For a perfect de Sitter space, where $H$ is constant, we have
\begin{equation}
x = \frac{k}{aH} \ \ \ \mbox{and} \ \ \ p = \frac{2\pi}{H},
\end{equation}
from which it follows that
\begin{equation}
\frac{p'}{p} = 0 \ \ \ \mbox{and} \ \ \ q = \frac{p'' - 3p'}{p} =
0.
\end{equation}
Therefore, from Eq.~(\ref{ps}), only the leading term survives and we obtain the
simple result
\begin{equation}
\mathcal{P}_\psi(k) = \left( \frac{H}{2\pi} \right)^2,
\end{equation}
and the spectral index is of course exactly flat, that is,
\begin{equation}
n_\psi(k) = 0.
\end{equation}
This is because $\mathcal{P}_\psi(k)$ depends only on $H$ during inflation, not
the detailed dynamics of the inflaton field $\phi$ due to the potential
$V(\phi)$. That is, the gravitational waves depend only on the energy scale
associated with the inflaton potential, which is one characteristic feature of
the tensor perturbations. So, even if $V(\phi)$ has some features, e.g. a linear
potential with a slope change \cite{cgs}, as long as $V(\phi)$ is very close to
flat, so that de Sitter space is a good approximation, we obtain a featureless,
nearly flat $\mathcal{P}_\psi(k)$, but nontrivial $\mathcal{P_R}(k)$.

\subsection*{Acknowledgements}

I am grateful to Ewan Stewart for many important comments on
earlier drafts. It is also my great pleasure to thank Jai-chan
Hwang for encouragement and useful suggestions, and Misao Sasaki
for insightful correspondences. This work was supported in part by
ARCSEC funded by the Korea Science and Engineering Foundation and
the Korean Ministry of Science, and Brain Korea 21.

\end{document}